\documentclass[pdflatex,sn-mathphys-num]{sn-jnl}

\usepackage{graphicx}%
\usepackage{multirow}%
\usepackage{amsmath,amssymb,amsfonts}%
\usepackage{amsthm}%
\usepackage{mathrsfs}%
\usepackage[title]{appendix}%
\usepackage{xcolor}%
\usepackage{textcomp}%
\usepackage{manyfoot}%
\usepackage{booktabs}%
\usepackage{algorithm}%
\usepackage{algorithmicx}%
\usepackage{algpseudocode}%
\usepackage{listings}%

\theoremstyle{thmstyleone}%
\theoremstyle{thmstyletwo}%

\theoremstyle{thmstylethree}%

\begin{document}

\title[Article Title]{The Generalized Second Law for Kerr-Taub-NUT Black Holes Perturbed by Test Fields}


\author[1]{\fnm{Koray} \sur{D\"{u}zta\c{s}}}\email{koray.duztas@okan.edu.tr}

\author[2,3]{\fnm{Ahmet Cem} \sur{Erdo\u{g}an}}\email{ahmet.erdogan@yeditepe.edu.tr}

\affil{$^1$Faculty of Engineering and Natural Sciences, Istanbul Okan University, Istanbul, T\"{u}rkiye}

\affil{$^2$ Department of Physics, Yeditepe University, Istanbul, T\"{u}rkiye}

\affil{ $^{3}$ Department of Physics, Bo\u{g}azi\c{c}i University, \.{I}stanbul, T\"{u}rkiye}

\abstract{We evaluate the variations in the area and the entropy of Kerr-Taub-NUT black holes perturbed by bosonic and fermionic test fields. We demonstrate that the critical frequency for a test field at which the variation of the area vanishes is strictly larger than the superradiance limit for Kerr-Taub-NUT black holes. Therefore both bosonic fields in the relevant range and fermionic fields which do not exhibit superradiant scattering can induce a decrease in the area of a Kerr-Taub-NUT black hole. However, the entropy is not directly proportional to the area for Kerr-Taub-NUT black holes. Only fermionic fields with frequencies below the superradiance limit can induce an entropy decrease. The simultaneous loss of entropy by the black hole and the environment identifies a generic violation of the Generalized Second Law (GSL) for fermionic perturbations that fail to satisfy the Null Energy Condition (NEC). For bosonic perturbations that satisfy the NEC, we evaluate the limiting case, where a nearly extremal black hole is perturbed by a scalar field with frequency slightly above the superradiance limit. Our analysis reveals that the minimum increase in the black hole entropy compensates for the entropy loss of the environment, when one assigns Von-Neumann entropy to the test field. The GSL remains valid for perturbations satisfying the NEC. }


\maketitle

\section{Introduction}
The Kerr-Taub-NUT (KTN) metric represents the most general stationary, axi-symmetric, Petrov Type-D  vacuum solution of the Einstein field equations. Generalizing the original Taub-NUT spacetime to include an angular momentum parameter, this solution was initially obtained by Demia\v{n}ski and Newman \cite{deminewman}. Adopting Kerr-like coordinates as presented by Miller \cite{miller}, the metric takes the following form: 
\begin{equation}
d s^2=\frac{1}{\Sigma}(\Delta -a^2\sin^2\theta)\,d t^2-\frac{2}{\Sigma}[A' \Delta -a(\Sigma +a A')\sin^2\theta]\,d t
d \phi-\frac{1}{\Sigma}[(\Sigma +a A')^2\sin^2\theta-{A'}^2\Delta]\, d \phi^2 -\frac{\Sigma}{\Delta}d r^2 -\Sigma \,d\theta^2
\label{ktnmetric1}
\end{equation}
In the KTN metric (\ref{ktnmetric1}), $\Sigma$, $\Delta$ and $A'$ are given by
\begin{eqnarray*}
&\Sigma& = r^2 +(\ell +a \cos \theta)^2 \\ 
&\Delta& = r^2-2Mr-\ell^2 + a^2 \nonumber \\ 
&A'& = a \sin^2\theta -2\ell\cos\theta 
\end{eqnarray*}
where  $M$, $-\ell$ and $a$ are the Schwarzschild (gravitational mass), NUT (gravitomagnetic charge) and Kerr (spin) parameters, respectively. When $a$ is zero, the metric reduces to the Taub-NUT metric, and when $\ell$ is zero, it reduces to the Kerr metric. 
The real and positive roots of the equation $\Delta=0$ correspond to the inner and outer event horizons of the KTN spacetime, with spatial locations
\begin{equation}
r_{\pm}= M \pm \sqrt{M^2+\ell^2-a^2}.
\label{eq:outereventhorizon}
\end{equation}
The term $\frac{\Sigma}{\Delta}$ is positive for both $r<r_-$ and $r_+<r$. These regions are designated NUT+ and NUT-. In the interval $r_-<r<r_+$, the term $\frac{\Sigma}{\Delta}$ is negative; therefore the r-coordinate and the t-coordinate exchange roles. This causes their spacelike and timelike natures to interchange. This interior region is referred to as the Taub region. In 1951, Taub solved the vacuum Einstein field equations under the assumption that the metric admits a three-parameter isometry group \cite{taub}. His solution contains two integration constants $M$ and $\ell$. Independently, Newman, Tamburino and Unti found a vacuum solution to the Einstein field equations that generalizes the Schwarzschild solution by introducing new arbitrary parameter $\ell$ in addition to the mass parameter $M$\cite{nut}. Misner subsequently demonstrated that the NUT solution includes the Taub solution in a limited region \cite{misner-taubnut}. The r-coordinate in the NUT solution corresponds to the t-coordinate in the Taub solution. Thus, the parameter $\ell$ is an inherent geometric property already present in Taub spacetime; the NUT charge is not added afterwards. 

Despite its intrinsic geometric origin, the physical meaning of the NUT charge remains controversial. Lynden-Bell and Nouri-Zonoz interpreted it as a gravitomagnetic monopole \cite{Lynden-Bell}. Henneaux and Teitelboim argued that the NUT charge is not a real source, but a duality in linearized gravity \cite{HT2005}. Al-Badawi and Halilsoy modeled a spacetime that has a Schwarzschild parameter and an externally powered twisting electromagnetic universe \cite{BH2006}. They concluded that the NUT parameter is related to the twisting parameter. Though the physical meaning of the NUT parameter remains controversial, the geometric and causal properties of the KTN spacetime have constituted an active area of research.

The KTN spacetime is not asymptotically flat i.e. it does not approach Minkowski spacetime at infinity.  Furthermore, the geometry possesses string-like singularities along the axis of symmetry ($\theta=0, \pi$), referred to as Misner strings. These represent torsion singularities, characterized by a singular metric determinant, rather than curvature or conical singularities \cite{torsionsingularity}. While these singularities can be removed by identifying the time coordinate $t$ with a period of $8\pi l$, this procedure introduces closed timelike curves. Consequently, the KTN spacetime is not globally hyperbolic as it does not admit a Cauchy hypersurface. Due to these peculiar geometric properties, the KTN metric is usually regarded as an analytic tool to probe the possible effects of gravito-magnetic monopoles, rather than a description of an astrophysical object. However, its observational verification remains accessible via its shadow, which is an essential tool for the direct detection of black holes \cite{blackholeobservation}. The fact that the NUT parameter introduces distinct deformations to the shadow geometry that diverge from the standard Kerr prediction \cite{ktnobservation}, renders the identification of KTN black holes possible. In addition to its observational, geometrical, and causal aspects, the thermodynamic properties of the KTN black hole are also of particular interest.

In the 1970s, the analogies between the black hole mechanics and thermodynamics were revealed \cite{bch}. Assuming the null energy condition (NEC), Hawking derived the area theorem which states that
\begin{equation}
\frac{d A_{\rm{BH}}}{d t}\geq 0
\label{areaentropy}
\end{equation} 
i.e. the area $A_{\rm{BH}}$ of the black hole cannot decrease under any classical process \cite{hawkingarea,hawkingarea2}. The apparent analogy between the second law of thermodynamics and the area theorem inspired Bekenstein to propose that black holes  possess entropy \cite{bekensteinentropy} proportional to their area $S \propto A$. Hawking later determined the proportionality constant, thereby establishing the relation between black hole entropy and horizon area \cite{hawking/4}.
\begin{equation}
S=\frac{kc^3}{G\hbar}\frac{A}{4}
\label{eq:entropygeneralwithunits}
\end{equation}
where $k$ is the Boltzmann constant, $c$ is the speed of light, $G$ is the gravitational constant and $\hbar$ is the reduced Planck constant. In this work we adapt natural units $G=c=\hbar=k=1$, in which the entropy area relation (\ref{eq:entropygeneralwithunits}) can be written as $S=A/4$.

In classical general relativity, a decrease in the area of  a black hole cannot be achieved by any classical process satisfying the NEC; therefore the area theorem holds. However, in the semiclassical regime, the area theorem may not remain valid due to quantum mechanical effects. The process of Hawking radiation which extracts  energy and angular momentum from a black hole, comprises a counter-example to the area theorem. As a solution to this problem Bekenstein suggested the generalized second law  \cite{bekensteinentropy,bekensteingsl1}, which states that the total entropy of a black hole and the exterior region never decreases. 
\begin{equation}
\Delta S_{\rm{BH}} + \Delta S_{\rm{ext}} \geq 0
\end{equation}
The validity of the GSL across all physically permissible processes would imply that black hole entropy is a genuine physical quantity. Under this paradigm, the correspondence between the laws of thermodynamics and black hole mechanics transcends mere mathematical analogy, establishing a unified thermodynamic reality. 

Bekenstein devised a classic thought experiment in which a spherical test body of mass $M$ (or energy $\delta M$) and radius $R$ is lowered near a black hole horizon and released. The absorption of the body increases the black hole?s surface area by a minimal amount, $\Delta A_{\text{min}} = 8\pi (\delta M)R$. To preserve the GSL, the resulting gain in black hole entropy must compensate for the loss of entropy in the exterior region. Bekenstein proposed a universal entropy bound for any localized system $S \le 2\pi (\delta M)R$, to ensure the validity of the GSL \cite{bekensteinbound}. However the universality of this bound remains controversial \cite{pagecomment,deutch}. Unruh and Wald argued that the bound is unnecessary because the buoyancy force exerted by Hawking radiation prevents the object from being lowered arbitrarily close to the horizon \cite{unruhwald1982,unruhwald1983}. However, for macroscopic bodies, this buoyancy effect is practically negligible, leading to counter-rebuttals by Bekenstein \cite{bekenbuo1,bekenbuo2}. Alternatively, maintaining a macroscopic object quasi-statically near the horizon requires work by an external agent, which itself generates entropy. When accounting for the entropy generated during this interaction process, the GSL holds without the need to employ Bekenstein?s bound \cite{gaowald}. In \cite{dgsl} we noted that, whether Bekenstein's bound is required to protect the GSL is logically distinct from the broader question of whether such a universal limit exists in nature.

Subsequently, the first mathematical proof of the GSL was given by  Frolov and Page in 1993, under certain assumptions  \cite{fp93}. Jacobson and Parentani proved the GSL for causal horizons which refer to boundaries of the past of time-like curves of infinite proper length in the future direction \cite{jacob}. Aron Wall  provided a review of the various attempts to prove the GSL in \cite{wall}. Wall also extended the proofs to rapidly changing quantum fields restricted to arbitrary slices of the horizon. \cite{wall2010,wall2012}. Since the outside entropy term develops an ultraviolet divergence at the horizon due to the entanglement entropy of the fields, he employed a renormalization scheme. The derivation is based on a non-trivial algebra of horizon operators which satisfy four axioms namely; determinism, ultralocality, local Lorentz symmetry and stability. The stability condition asserts that the fields on each horizon generator have positive energy. The assumptions  render the proofs in applicable to KTN space-time and fermionic test fields which violate  determinism and the energy condition, respectively. We have recently derived that the GSL is generically violated in the interaction of Kerr black holes with test fermionic fields that do not satisfy the NEC, while it remains valid for bosonic perturbations \cite{dgsl}. The absorption of fermionic test fields with frequencies below the superradiance limit leads the area of a Kerr black hole to decrease. The simultaneous loss of entropy of  the black hole and the environment represents a generic violation of the GSL. The GSL is contingent upon the validity of the NEC as well as the area theorem.

The Kerr-Taub-NUT spacetime is not asymptotically flat; therefore, the definition of global conserved charges, such as mass, is not unique, depending on the choice of asymptotic structure, and its thermodynamic interpretation remains subtle. In the Euclidean sector, one disregards the Misner strings by imposing periodic time; thus, a new parameter related to the NUT charge cannot be added to the first law of the black hole dynamics.  This implies that equation (\ref{eq:entropygeneralwithunits}) is not satisfied i.e. the entropy of a KTN black hole is not directly proportional to its area \cite{hennigar2019,mann2000}. In the Lorentzian sector, one considers the Misner strings. By adding the new parameter, one can attempt to determine the entropy of the KTN black hole \cite{durka2022,bordo2019,awad2022,yang2023}. Frodden and Hidalgo evaluated the variation in the entropy of a KTN black hole and established the first law for the KTN spacetime \cite{frodden2022}.

Previously we have evaluated the possibility to destroy the event horizon of a KTN black hole, and derived that bosonic fields that satisfy the NEC cannot destroy the event horizon provided that one incorporates the second order variations in the limiting cases \cite{dktn,dktn2}. However the absorption of fermionic fields with frequencies below the superradiance limit leads to a generic destruction of the event horizon, in accord with our derivations for the Kerr case \cite{generic,fermionic}. In this work, we investigate the validity of the area theorem and the GSL for Kerr-Taub-NUT black holes interacting with massless fermionic and bosonic fields. For that purpose, we examine the variation in the area of the black hole and the entropy of the exterior region when a test field with frequency $\omega$ is absorbed by the black hole. We analyse fermionic and bosonic fields separately as the latter would be subject to the superradiance condition. We assign Von-Neumann entropy to test fields to evaluate the validity of the GSL.

\section{The variation in the area}
\label{sec:area}
In this section, we calculate the variation in the area of a KTN black hole perturbed by bosonic or fermionic test fields to check whether it can be negative for particular modes of the test field. The area of the Kerr-Taub-NUT black hole is
\begin{equation}
A_{\rm{BH}}= 4\pi (r_+^2+a^2+\ell^2)
\label{eq:area}
\end{equation} 
where $r_+$ is the spatial radius event horizon of the KTN black hole given in equation \ref{eq:outereventhorizon}, $a=\frac{J}{M}$ is angular momentum per unit mass and $\ell$ is the NUT charge. The area (\ref{eq:area}) can be expressed in the form:
\begin{equation}
A_{\rm{BH}}=8\pi \left[ 	M^2 + \sqrt{M^4-J^2+M^2\ell^2}+ \ell^2  \right].
\end{equation}
The variation of the area of the black hole is 
\begin{equation}
\delta A_{\rm{BH}}= 8\pi \left[
2M \, \delta M + \frac{4M^3+2M\ell^2}{2\sqrt{M^4-J^2+M^2\ell^2}} \, \delta M  - \frac{2 J}{2\sqrt{M^4-J^2+M^2\ell^2}}\, \delta J   
\right]
\end{equation}

A test field or test particle that induces a change in the black hole area carries neither NUT charge nor any quantity associated with the NUT parameter, i.e. $\delta l =0$.

For a stationary, axisymmetric vacuum spacetime, one can obtain the relation between the energy and the angular momentum parameters of test field.
\begin{equation}
\delta J =\frac{m}{\omega}\, \delta M
\label{eq:angular momentum and mass}
\end{equation} 
where $m$ is the azimuthal quantum number and $\omega$ is the mode frequency of the field.
One finds that the ratio of the change in the angular momentum of a test field and the change in its energy for a fixed $m$ increases as the mode frequency $\omega$ decreases. We determine the critical frequency for which the variation of the area of the black hole vanishes. By using (\ref{eq:angular momentum and mass}), we obtain
\begin{equation}
\delta A_{\rm{BH}}= 8\pi 
\left[
2M + \frac{2M^3+M\ell^2}{\sqrt{M^4-J^2+M^2\ell^2}} - \frac{Jm}{\omega\sqrt{M^4-J^2+M^2\ell^2}} 
\right] \, \delta M  =0
\label{deltaa1}
\end{equation}
Thus,
\begin{equation}
\omega_{\rm{crit}} = \frac{ma}{2M r_+ + \ell^2}=\frac{ma}{r_+^2+a^2}
\label{omegacrit}
\end{equation}
This implies that the area of a KTN black hole decreases when the frequency $\omega$ is less than  $\omega_{\rm{crit}}$. Recently we derived that the critical frequency for which the variation in the area vanishes, coincides with the superradiance limit for Kerr black holes \cite{dgsl}. However, there is a subtle difference in the case of KTN black holes. The angular velocity of the event horizon is given by
\begin{equation}
\Omega_{\rm{EH}}=\frac{a}{r_+^2+a^2+\ell^2}=\frac{a}{2(Mr_++\ell^2)}.
\label{eq:omegaeh}
\end{equation}
It has been verified  that superradiance occurs for bosonic fields that satisfy the NEC, whose frequency is less than the superradiance limit \cite{bini,lee}
\begin{equation}
\omega_{\rm{sl}}=m\Omega_{\rm{EH}}.
\end{equation}
The superradiance limit for KTN black holes is strictly less than the critical frequency (\ref{omegacrit}) for which $\delta A=0$. Therefore bosonic fields with frequencies in the range $\Omega_{\rm{sl}} < \omega <\Omega_{\rm{crit}}$ can be absorbed by a KTN black hole and induce a decrease in the area of the black hole. The superradiance limit does not apply to fermionic fields, therefore the range of frequencies that lead the area of a KTN black hole to decrease does not have a lower bound. $\delta A$ becomes negative if the frequency of a fermionic field is below the critical value $\Omega_{\rm{crit}}$, without a lower bound.

The main assumption in the proof of the area theorem is the NEC. Therefore fermionic fields which can circumvent the NEC, are not restricted by the area theorem. Naively, one would expect the absorption of fermionic fields with low frequencies to induce a decrease in the area of a KTN black hole, in accord with our recent analysis on Kerr black holes \cite{dgsl}. However, the result that bosonic fields can also lead the area to decrease appears as a violation of the area theorem which requires a justification. 

There is a second assumption in the proof of the area theorem, which asserts that the horizon generators are complete. The area of the event horizon is non-decreasing if the expansion $\theta$ of the horizon generators is non-negative. If one assumes that the horizon generators are complete in the sense that they never develop focal points and they can be extended to arbitrary values of the affine parameters, one can prove the non-negativity of the expansion via Raychaudri equation. However Taub-NUT and Kerr-Taub-NUT space-times exhibit a characteristic property known as imprisoned incompleteness. Incomplete time-like and null geodesics are totally imprisoned in compact neighbourhoods of the event horizon. These space-times are singular according to the definition of Penrose and Hawking based on geodesic incompleteness, although the null geodesics do not develop a focal point. The imprisoned geodesics correspond to incomplete null geodesics spiralling indefinitely around $\tau=0$, in Misner space-time which is a $(1+1)$ dimensional toy model to study the Taub-NUT space-time (See e.g. \cite{hawkingellis}). There exist an incomplete, closed null geodesic at $\tau=0$ which correspond to incomplete closed null geodesics on the horizon of Taub-NUT and KTN black holes.  Therefore KTN black holes do not satisfy the assumptions of the area theorem.

Recently we have calculated the second order variation of the area ($\delta^2 A$) of a nearly extremal KTN black hole \cite{dktn2}. The second order variation involves $(\delta M)^2$, $(\delta J)^2$, $(\delta M)(\delta J)$, $\delta^2 M$ and $\delta^2 J$ terms. We showed that $\delta^2 M$ and $\delta^2 J$ terms can be eliminated using the Sorce-Wald condition \cite{sorcewald} 
\begin{equation}
\delta^2 M-\Omega \delta^2 J \geq  \frac{\kappa}{8\pi}\delta^2 A .
\end{equation}
Then, the second order variation can be expressed in terms of  $(\delta J)^2$ or $(\delta M)^2$
\begin{equation}
\delta^2 A =-\frac{8\pi}{M^8 \epsilon^3}\left \{  M^6 \frac{\epsilon^2}{2} \right\}(\delta J)^2=-\frac{4\pi}{M^2 \epsilon}\delta J^2=-\frac{4\pi}{M^2 \epsilon}\frac{m^2}{\omega^2}\delta M^2 .
\end{equation}
If we incorporate the effect of the second order variations in (\ref{deltaa1}), we need to solve a tedious quadratic equation for the critical frequency $\omega_{\rm{crit}}$. In that case $\omega_{\rm{crit}}$ undergoes a slight increase. (We assume that $\omega$ is positive so that the contribution of the test field to the angular momentum is positive.) The condition $\omega_{\rm{crit}}> \omega_{\rm{sl}}$ trivially maintains its validity, therefore the conclusion about bosonic and fermionic fields remain valid. 
\section{The entropy variation}
\label{sec:entropy}
In this section, we evaluate the variation in the entropy of a KTN black hole perturbed by bosonic and fermionic test fields, adopting the entropy expression given in \cite{frodden2022}.
\begin{equation}
{\delta} S  =   \frac{1}{T} \left( \delta M- \Omega \delta J\right)=\frac{2\pi }{\sqrt{M^2-a^2+\ell^2}}
 \left(  \big(2 M (M+\sqrt{M^2-a^2+\ell^2})- a^2+2\ell^2\big)\,\delta M- M a \, \delta a \right)\,
\label{deltas1}
\end{equation}
For a KTN black hole, the entropy is not merely given by the area multiplied by a proportionality constant. Although the expression for the entropy cannot be analytically integrated, the relation between the variation in the area and the entropy of the KTN black hole can be expressed as
\begin{equation}
\delta S= \frac{\delta A}{4}+\frac{2\pi}{\sqrt{M^2-a^2+\ell^2}}\, \ell^2 \, \delta M
\label{entropyktn1}
\end{equation}
Manifestly one obtains the linear dependence in the form (\ref{areaentropy}) in the limit $\ell \to 0$.
Imposing $a=\frac{J}{M}$ and $\delta J=\frac{m}{\omega}\, \delta M$, we obtain
\begin{equation}
\delta a= \frac{1}{M}\left( \frac{m}{\omega}- a\right)\delta M
\end{equation}
We substitute this expression in (\ref{deltas1}), 
\begin{equation}
\label{eq:entropy}
{\delta} S =\frac{2\pi }{\sqrt{M^2-a^2+\ell^2}}
\left(  2 M (M+\sqrt{M^2-a^2+\ell^2})+2\ell^2 - \frac{ma}{\omega} \right)\,\,\delta M.
\end{equation}
The equation (\ref{eq:entropy}) implies that $\delta S=0$ for $\omega=\omega_{\rm{sl}}$. The absorption of the modes with frequencies $\omega<\omega_{\rm{sl}}$, would result in a negative value for $\delta S$. Therefore, only fermionic perturbations can induce a decrease in the entropy of the KTN black hole. Simultaneously, the environment also loses entropy due to the absorption of the test field, which marks a generic violation of the GSL.
\section{Limiting cases for bosonic fields}
\label{sec:bosonic}
The analysis in section \ref{sec:entropy} implies that bosonic fields satisfying the NEC cannot lead to a generic violation of the GSL. Therefore we evaluate the limiting cases for bosonic fields to check whether the minimum increase in the entropy of the black hole can compensate for the decrease in the entropy of the environment, as the test field is absorbed by the KTN black hole. Relevant analyses were previously carried out for Reissner-Nordstr\"{o}m \cite{hod} and Kerr black holes \cite{dgsl}.
In order to minimize the increase in entropy,  we consider a nearly extremal KTN black hole perturbed by a scalar field with frequency slightly above the superradiance limit. We parametrize a nearly extremal black hole as:
\begin{equation}
M^2-a^2+\ell^2=M^2\epsilon^2
\label{param1}
\end{equation}
We introduce the dimensionless parameters:
\begin{equation}
\varepsilon^2= 1-\frac{a^2}{M^2}+\frac{\ell^2}{M^2}=1-\alpha^2+\beta^2
\end{equation}
where we defined $\alpha=\frac{a}{M}$ and $\beta=\frac{\ell}{M}$, as we did in \cite{dktn,dktn2}.
For a KTN black hole parametrized as (\ref{param1}), the spatial radius of the event horizon is given by
\begin{equation}
r_+= M(1+\varepsilon).
\end{equation}
Similarly the expression for the angular velocity of the KTN black hole  given  \ref{eq:omegaeh}, takes the form
\begin{equation}
\Omega_{\rm{EH}}=\frac{\alpha}{2M (\varepsilon+\varepsilon^2+\alpha^2 )}=\frac{\alpha}{2M (\varepsilon+1+\beta^2 )}.
\label{eq:omegaeffective}
\end{equation}
Thus, we get
\begin{equation}
\omega_{\rm{sl}}=\frac{m\alpha}{2M (\varepsilon+\varepsilon^2+\alpha^2 )}=\frac{m\alpha}{2M (\varepsilon+1+\beta^2 )}
\end{equation}
for the limiting frequency for superradiance.
The surface gravity $\kappa$ of a KTN black hole is given by
\begin{equation}
\kappa=\frac{r_+-r_-}{2(r_+^2+a^2+\ell^2)}
\label{eq:surfacegravity}
\end{equation}
For a nearly extremal KTN black hole as parametrized as in (\ref{param1}), we obtain
\begin{equation}
\kappa=\frac{\varepsilon}{2M(\varepsilon+\beta^2+1)}
\end{equation}
Hawking temperature is proportional to the surface gravity, which takes the form
\begin{equation}
T=\frac{\kappa}{2\pi}=\frac{\epsilon}{4\pi M(\varepsilon+\beta^2+1)}.
\label{eq:temperature}
\end{equation}
We adjust the frequency of the bosonic test field to be slightly larger than the superradiance limit to minimize the area increase and to ensure that the test field is absorbed by the black hole.
\begin{equation}
\label{eq:testomega}
\omega=\omega_{\rm{sl}}(1+\epsilon)= m\,\Omega_{\rm{EH}}(1+\epsilon)
\end{equation}
This choice of the frequency implies:
\begin{equation}
\frac{ma}{\omega}=\frac{2M^2}{1+\epsilon}(\epsilon +1+\beta^2)=2M^2(1+\beta^2 - \epsilon \beta^2 +\epsilon^2 \beta^2)
\label{amomega}
\end{equation}
Plugging (\ref{amomega}) into (\ref{eq:entropy}), we derive 
\begin{eqnarray}
{\delta} S &=& \frac{2\pi }{M\epsilon}
\left(  2M^2\epsilon + 2\ell^2 \epsilon - 2 \ell^2 \epsilon^2 \right)\,\,\delta M \nonumber \\
&=& \frac{ 4 \pi}{M} \left[ M^2 + \ell^2 (1-\epsilon)\right]\,\delta M
\label{deltas2}
\end{eqnarray}
We choose  ${\delta} M =M\eta$, for the contribution of the test field to the mass parameter, where $\eta \ll 1$ in accord with the test field approximation. Note that the small parameters describing the closeness to extremality ($\epsilon$) and the energy of the test field ($\eta$) need not be equal. Substituting $\delta M$ in (\ref{deltas2}), we derive the variation in the entropy of the KTN black hole.
\begin{equation}
\delta S= 4\pi \eta \left[ M^2 + \ell^2 (1-\epsilon)\right]
\label{deltas3}
\end{equation}
The change in the entropy of a nearly extremal KTN black hole is first order in $\eta$ when we consider the limit $\epsilon \to 0$, i.e. when a test field with frequency arbitrarily close to the superradiance limit is absorbed.
Next, we calculate the change in the entropy of the environment. The environment loses entropy as the test field is absorbed by the KTN black hole. We check if the loss of the entropy by the environment can be compensated by the increase in the entropy of the KTN black hole derived in (\ref{deltas3}). Our purpose is to examine the generalized second law within the semiclassical test-field framework by comparing the change in the black-hole entropy with the entropy of the test field. We assign von Neumann entropy to the bosonic test field which equals the entropy of an identical test field emitted by the black hole via Hawking radiation. The test field sent in from infinity is not entangled with the internal states of the black hole. In this analysis we ignore the actual Hawking radiation, its associated entanglement, and quantum correlations. The von Neumann entropy is given by 
\begin{equation}
\delta S = (N+1)\ln (1+N)-N \ln N,
\label{vonneumann}
\end{equation}
where $N$ is the average number of particles in Hawking radiation,
\begin{equation}
N_{\omega lm}=\frac{\Gamma_{lm}(\omega)}{\exp [(\omega - m\Omega)/T] \pm 1},
\end{equation} 
where $\Gamma_{lm}$ is greybody factor,  $l$ is the orbital quantum number, $m$ is the azimuthal quantum number, $\omega$ is the mode frequency of the test field, $\Omega$ is the angular velocity of the event horizon, and $T$ is the Hawking temperature. The $\pm$ sign in the denominator corresponds to fermionic and bosonic fields, respectively. The greybody factor is the absorption probability that a field in a given incoming wave mode will be absorbed by a black hole. If $\Gamma^s_{lm}=0$, the field is entirely reflected back to infinity.  If $\Gamma^s_{lm}$ is negative, the field is reflected back with a larger amplitude, which corresponds to superradiance. (In \cite{superrad}, it was pointed out that the conventional term absorption probability is misleading as $\Gamma$ can attain negative values.)  

The greybody factors for Kerr black holes for bosonic and fermionic perturbations were derived by Page in a seminal work \cite{page}. The corresponding factors for KTN black holes are given by \cite{lee},
\begin{equation}
\Gamma^s_{lm}  = \bigg( 2^l l! \frac{ (l-s)! (l+s)! }{ (2l)! (2l+1)! } \bigg)^2  \prod_{n=1}^l \bigg[ 1 + \bigg( \frac{1}{n} \frac{(\omega-m\Omega)}{2 \pi T} \bigg)^2 \bigg] \frac{(\omega-m\Omega)}{\pi T} (A_{\rm{BH}} T \omega)^{2l+1} , \, 2l \in \rm{even}.
\end{equation}
We choose to perturb the KTN black hole by a scalar field with $s=0$. The highest absorption probability for $s=0$ pertains to the modes with $l=0$. However this choice entails that $m=0$, which does not contribute to the angular momentum parameter. Therefore we choose  $l=1$ and $m=1$ for the scalar field.  Using (\ref{eq:omegaeffective}) for $\Omega$, (\ref{eq:testomega}) for $\omega$, (\ref{eq:temperature}) for $T$ and (\ref{eq:area}) for $A_{\rm{BH}}$, we obtain
\begin{equation}
\Gamma^0_{11}  = \frac{1}{36}  \bigg[ 1 + \alpha^2  \bigg]  (2\alpha) \bigg[ \frac{\alpha \epsilon (1+\epsilon)}{\epsilon+1+\beta^2} \bigg]^3
\end{equation}
Note that the lowest order contribution to the greybody factor is $\epsilon^3$. The average number of particles is given by
\begin{equation}
N=\frac{\Gamma^0_{11}}{\exp [2\pi\alpha] - 1} \sim \frac{\epsilon^3}{\exp [2\pi\alpha] - 1}
\label{Nepsiloncube}
\end{equation}
In \cite{hawkingzero} we showed that the thermal radiation  derived by Hawking can be smoothly extended to $T \to 0$ limit, which is corresponds to $\epsilon \to 0$ in (\ref{eq:temperature}). Using the expression (\ref{Nepsiloncube}) for $N$,
\begin{equation}
\lim_{\epsilon \to 0} N= 0.
\end{equation}
Using (\ref{vonneumann}), the decrease in the entropy of the outer region will approach zero when $\epsilon \to 0$.
\begin{equation}
\lim_{\epsilon \to 0} \delta S= 0
\label{eq:the change in the entropy of the outer region}
\end{equation}
Comparing the increase of the entropy of the KTN black hole (\ref{deltas3}) and the decrease in the entropy of the environment  (\ref{eq:the change in the entropy of the outer region}), one concludes that the variation in the total entropy of the KTN spacetime  is positive. Therefore, the generalized second law remains valid for bosonic fields, satisfying the NEC.

\section{Conclusions}
The GSL is regarded as a reliable assumptions in black hole physics and its validity underlies the physical interpretation of black hole entropy. Recently we have demonstrated that its validity is contingent upon the validity of the NEC, in the interaction of Kerr black holes with test fields \cite{dgsl}. The critical frequency below which the area of a Kerr black hole would decrease coincides with the superradiance limit. Therefore the absorption of fermionic fields with frequencies below the critical limit leads to a generic violation of the GSL for Kerr black holes. 

In the semiclassical limit, the perturbation of the metric is small compared to the unpertubed metric. The zeroth-order perturbation corresponds to the classical metric. For the zeroth-order perturbation, Wall considered three possible cases for the classical evolution of the horizon: a classically growing horizon, a classically stationary horizon, and a horizon that grows classically up to a certain time and subsequently becomes stationary in \cite{wall2012}. He argued that, in the classically stationary case, quantum effects may cause the black-hole area to decrease which could lead to a violation of the GSL. He concluded that any  violation of the GSL must originate from quantum effects.

In section \ref{sec:area} we derived that the critical frequency for KTN black holes is strictly larger than the superradiance limit, unlike the Kerr case. There exists a range of frequencies for bosonic fields for which their absorption is permitted and the variation in the area becomes negative. Contrary to the statement in \cite{wall2012}, the decrease of the area is not caused by quantum effects. The area theorem can be violated by bosonic fields that satisfy the NEC as well as fermionic fields. We argued that the area theorem does not apply to KTN black holes since there exist incomplete, closed timelike curves on the horizon. However, the  entropy of a KTN black hole is not merely proportional to its area, as it incorporates contributions unique to the NUT-charged background, which is manifest in Equation (\ref{entropyktn1}). In section \ref{sec:entropy} we demonstrated that only the absorption of fermionic fields with frequencies below the superradiance limit can induce a decrease in the entropy. Simultaneously, the exterior region loses entropy which leads to a generic violation of the GSL. For bosonic fields, the variation of the entropy of a KTN black hole remains positive, since the absorption of the modes with frequencies below the superradiance limit is not permitted. In section \ref{sec:bosonic}, we evaluated the limiting case for bosonic fields, where a nearly extremal KTN black hole is perturbed by a scalar test field with frequency slightly above the superradiance limit. We assigned Von-Neumann entropy to the test field and calculated the entropies gained by the KTN black hole and lost by the exterior region. We demonstrated that the GSL remains valid for perturbations satisfying the NEC. 
 
In this work and the relevant study on Kerr black holes \cite{dgsl} we restricted the entropy of the exterior region to the entropy of the test field which is not entangled with the interior states of the black hole. We ignored the Hawking radiation emitted by the black hole, its associated entanglement entropy and non-local quantum correlations which are central in the modern formulation of the GSL. We derived that the absorption of the fermionic fields with frequencies below the superradiance limit leads to a generic violation of the GSL. However, the black hole will eventually evaporate and  the accumulating entropy of the emitted radiation and its associated quantum entanglement will contribute significantly to the exterior region. This accumulated quantum entropy can restore and guarantee the validity of the GSL over the full evaporation timescale of ($\sim 10^{70}$) years.

%
%







\end{document}